\documentclass[aps,prd,twocolumn,flushleft,amsmath,amssymb,nofootinbib]{revtex4}
\usepackage{subfig}
\usepackage{graphicx}
\usepackage{dcolumn} 
\usepackage{bm}      
\usepackage[active]{srcltx} 
\usepackage{hyperref} 
\usepackage{amsmath}
\usepackage{amssymb}
\usepackage{stmaryrd}
\usepackage{amsfonts}
\usepackage{mathrsfs}

\newcommand{\be}{\begin{equation}}
\newcommand{\ee}{\end{equation}}
\newcommand{\ba}{\begin{eqnarray}}
\newcommand{\ea}{\end{eqnarray}}
\newcommand{\Tr}{\mathrm{Tr}} 
\newcommand{\grav}{\mathrm{gr}} 
\newcommand{\sca}{\mathrm{sc}} 
\newcommand{\kin}{\mathrm{kin}} 
\newcommand{\SU}{\mathscr{S}\mathscr{U}} 
\newcommand{\hil}{\mathcal{H}} 
\newcommand{\R}{\mathcal {R}} 
\newcommand{\kt}{{\tilde{K}}} 

\def\nn{\nonumber}

\begin{document}

\title{Extension of loop quantum gravity to $f(R)$ theories}

\author{Xiangdong Zhang\footnote{zhangxiangdong@mail.bnu.edu.cn} and Yongge Ma\footnote{mayg@bnu.edu.cn}}
\affiliation{Department of Physics, Beijing Normal University, Beijing 100875, China}

\begin{abstract}
The 4-dimensional metric $f(\R)$ theories of gravity are cast into
connection-dynamical formalism with real $\SU(2)$-connections as
configuration variables. Through this formalism, the classical metric $f(\R)$ theories are quantized by extending the
loop quantization scheme of general relativity. Our results imply that the
non-perturbative quantization procedure of loop quantum gravity
is valid not only for general relativity but also for a rather
general class of 4-dimensional metric theories of gravity.\\

\end{abstract}

\maketitle

In recent twenty years, loop quantum gravity(LQG), a background
independent approach to quantize general relativity(GR), has been
widely investigated\cite{Ro04,Th07,As04,Ma07}. It is remarkable
that, as a non-renormalizable theory, GR can be non-perturbatively
quantized by the loop quantization procedure\cite{Le10}. This
background-independent quantization relies on the key observation
that classical GR can be cast into the connection-dynamical
formalism with the structure group of $SU(2)$\cite{As86,Ba}. Thus it
is interesting to see whether GR is a unique relativistic theory
with the connection-dynamic character. Especially, modified gravity
theories have received increasingly attention recently due to
motivations coming form cosmology and astrophysics. A series of
independent observations, including type Ia supernova, weak lens,
cosmic microwave background anisotropy, baryon oscillation, etc,
implied that our universe is currently undergoing a period of
accelerated expansion\cite{01}. This result conflicts with the
prediction of GR and has carried the "dark energy" problem. Hence it
is reasonable to consider the possibility that GR is not a valid
theory of gravity on a cosmological scale. Since it was found that a
small modification of the Einstein-Hilbert action by adding an
inverse term of curvature scalar $\R$ would lead to current
acceleration of our universe, a large variety of models of $f(\R)$
modified gravity have been proposed\cite{So}. Moreover, some models
of $f(\R)$ gravity may account for the "dark matter" problem, which
was revealed by the observed rotation curve of galaxy
clusters\cite{So}. Historically, Einstein's GR is the simplest
relativistic theory of gravity with correct Newtonian limit. It is
worth pursuing all alternatives, which provide a high chance to new
physics. Recall that the precession of Mercury's orbit was at first
attributed to some unobserved planet orbiting in side Mercury's
orbit, but was actually explained only after the passage from
Newtonian gravity to GR.

Given the strong motivations to $f(\R)$ gravity, it is desirable to
study such kind of theories at fundamental quantum level. For metric
$f(\R)$ theories, gravity is still geometry as GR. The differences
between them are just reflected in dynamical equations. Hence, a
background-independent and non-perturbative quantization for $f(\R)$
gravity is preferable. In this letter, we derive the connection-dynamical formulation of $f(\R)$ gravity by canonical
transformations from it's geometrical dynamics. The latter was
realized by introducing a non-minimally coupled scalar field to
replace the original $f(\R)$ action and then doing Hamiltonian analysis.
The canonical variables of our Hamiltonian formalism of $f(\R)$
gravity consist of $\SU(2)$-connection $A_a^i$ and it's conjugate
momentum $E^a_i$, as well as the scalar field $\phi$ and it's
momentum $\pi$. The Gaussian, diffeomorphism and Hamiltonian
constraints are also obtained, and they comprise a first-class
system. Loop quantization procedure is then naturally employed to
quantize $f(\R)$ gravity. The rigorous kinematical Hilbert space
structure of LQG is extended to loop quantum $f(\R)$ gravity by
adding a polymer-like quantum scalar field. As in LQG, the Gaussian
and diffeomorphism constraints can be solved at quantum level, and
the Hamiltonian constraint is promoted to a well-defined operator.
We use Greek alphabet for spacetime in indices, Latin alphabet
$a,b,c...$ for spatial indices and $i,j,k...$ for internal indices.

The original action of $f(R)$ theories reads: \ba S[g]=\frac12\int
d^4x\sqrt{-g}f(\mathcal {R}) \label{action0} \ea where $f$ is a
general function of $\R$, and we set $8\pi G=1$. By introducing an
independent variable $s$ and a Lagrange multiplier $\phi$, an
equivalent action is proposed as\cite{Na,Na01}: \ba
S[g,\phi,s]&=&\frac12\int d^4x\sqrt{-g}(f(s)-\phi(s-\mathcal
{R})).\label{action} \ea The variation of (\ref{action}) with
respect to $s$ yields $\phi=df(s)/{ds}\equiv f'(s)$.
Assuming $s$ could be resolved from the above equation, action
(\ref{action}) is reduced to \ba S[g,\phi] &=&\frac12\int
d^4x\sqrt{-g}(\phi \mathcal {R}-V(\phi))\equiv\int d^4x \mathcal {L}
\label{action1} \ea where $V(\phi)\equiv\phi s-f(s)$. It is easy to
see that the variations of (\ref{action1}) give the equations of
motion equivalent to that  from action (\ref{action0}). The virtue
of (\ref{action1}) is that it admits a treatable Hamiltonian
analysis\cite{Na}. By doing 3+1 decomposition and Legendre
transformation: \ba p^{ab}&=&\frac{\partial\mathcal
{L}}{\partial\dot{h}_{ab}}=\frac{\sqrt{h}}{2}[\phi(K^{ab}-Kh^{ab})-\frac{h^{ab}}{N}(\dot{\phi}-N^c\partial_c\phi)], \label{04} \nn\\
\pi&=&\frac{\partial\mathcal {L}}{\partial\dot{\phi}}=-\sqrt{h}K,
\ea where $h_{ab}$ and $K_{ab}$ are the induced 3-metric and the
extrinsic curvature of the spatial hypersurface $\Sigma$
respectively, and $K\equiv K^a_a$, the Hamiltonian of $f(R)$ gravity
can be derived as a liner combination of constraints as $\hil_{total}=\int_\Sigma N^aV_a+NH$, where $N$ and $N^a$ are
the lapse function and shift vector respectively, and the
diffeomorphism and Hamiltonian constraints read \ba V_a
&=&-2D^b(p_{ab})+\pi\partial_a\phi,\label{constraint1}\\
H
&=&\frac2{\sqrt{h}}(\frac{p_{ab}p^{ab}-\frac13p^2}{\phi}+\frac16\phi\pi^2-\frac13p\pi)\nn\\
&+&\frac12\sqrt{h}(V(\phi)-\phi R+2D_aD^a\phi).\label{constraint2}
\ea The symplectic structure is given by \ba
\{h_{ab}(x),p^{cd}(y)\}&=&\delta^{(c}_a\delta^{d)}_b\delta^3(x,y),\nn\\
\{\phi(x),\pi(y)\}&=&\delta^3(x,y). \label{poission}\ea
Straightforward calculations show that the constraints
(\ref{constraint1}) and (\ref{constraint2}) comprise a first-class
system similar to GR\cite{Na01}. Although the above Hamiltonian
analysis is started with the action (\ref{action1}) where a
non-minimally coupled scalar field is introduced, one can check
that the resulted Hamiltonian formalism is equivalent to the Lagrangian
formalism\cite{Zh10}.

Recall that the non-perturbative loop quantization of GR was based
on it's connection-dynamic formalism. It is very interesting to
study whether the previous geometric dynamics of $f(\R)$ modified
gravity also has a connection-dynamic correspondence. To this aim,
we first introduce the following canonical transformation on the
phase space of $f(\R)$ theories. Let \ba\tilde{K}^{ab}\equiv\phi
K^{ab}+\frac{h^{ab}}{2N}(\dot{\phi}-N^c\partial_c\phi),\ea and
$E^a_i\equiv\sqrt{h}e^a_i$ where $e^a_i$ is the triad s.t.
$h_{ab}e^a_ie^b_j=\delta_{ij}$. Then we get \ba p^{ab}
&=&\frac{1}{2}(\tilde{K}^a_iE^{bi}-\frac1h\tilde{K}^i_cE^c_iE^a_jE^b_j), \nn\\
\pi&=&-\frac{\sqrt{h}}{\phi}(\tilde{K}^c_c-\frac{3}{2N}(\dot{\phi}-N^c\partial_c\phi)),
\ea where $\tilde{K}^a_i\equiv\tilde{K}^{ab}e_b^i$. By the
symplectic structure (\ref{poission}) we obtain the following
Poisson brackets: \ba
\{E^a_j(x),E^b_k(y)\}=\{\tilde{K}_a^j(x),\tilde{K}_b^k(y)\}=0,\nn\\
\{\tilde{K}^j_a(x),E_k^b(y)\}=\delta^b_a\delta^j_k\delta(x,y). \ea
Thus the transformation from conjugate pairs $(h_{ab},p^{cd})$ to
$(E^a_i,\tilde{K}^j_b)$ is canonical. Note that since
$\tilde{K}^{ab}=\tilde{K}^{ba}$, we have an additional constraint:
\be G_{jk}\equiv\tilde{K}_{a[j}E^a_{k]}=0. \label{gaussian}\ee So we
can make a second canonical transformation by defining: \be
A^i_a=\Gamma^i_a+\gamma\tilde{K}^i_a, \ee where $\Gamma^i_a$ is the
spin connection determined by $E^a_i$ and $\gamma$ is a nonzero real
number, since the Poisson brackets among the new variables read \ba
\{A^j_a(x),E_k^b(y)\}&=&\gamma\delta^b_a\delta^j_k\delta(x,y),\nn\\
\{A_a^i(x),A_b^j(y)\}&=&0. \ea Now, the phase space consists of
conjugate pairs $(A_a^i,E^b_j)$ and $(\phi,\pi)$. Combining
Eq.(\ref{gaussian}) with the compatibility condition:
$\partial_aE^a_i+\epsilon_{ijk}\Gamma^j_aE^{ak}=0$, we obtain the
standard Gaussian constraint \ba \mathcal
{G}_i=\mathscr{D}_aE^a_i\equiv\partial_aE^a_i+\epsilon_{ijk}A^j_aE^{ak},
\label{GC}\ea which justifies $A^i_a$ as an $\SU(2)$-connection.
Note that, had we let $\gamma=\pm i$, the (anti-)self-dual complex
connection formalism would be obtained. The original diffeomorphism
constraint can be expressed in terms of new variables up to Gaussian
constraint as \ba V_a &=&\frac1\gamma
F^i_{ab}E^b_i+\pi\partial_a\phi, \ea where
$F^i_{ab}\equiv2\partial_{[a}A^i_{b]}+\epsilon^i_{kl}A_a^kA_b^l$ is
the curvature of $A_a^i$. The original Hamiltonian constraint can be
written up to Gaussian constraint as \ba
H&=&\frac{\phi}{2}[F^j_{ab}-(\gamma^2+\frac{1}{\phi^2})\varepsilon_{jmn}\tilde{K}^m_a\tilde{K}^n_b]\frac{\varepsilon_{jkl}
E^a_kE^b_l}{\sqrt{h}}\nn\\
&+&\frac12(\frac2{3\phi}\frac{(\tilde{K}^i_aE^a_i)^2}{\sqrt{h}}+
\frac43\frac{(\tilde{K}^i_aE^a_i)\pi}{\sqrt{h}}+\frac23\frac{\pi^2\phi}{\sqrt{h}} \nn\\
&+&\sqrt{h}V(\phi))+\sqrt{h}D_aD^a\phi.\label{hamilton} \ea It is
easy to check that the smeared Gaussian constraint, $\mathcal
{G}(\Lambda):=\int_\Sigma d^3x\Lambda^i(x)G_i(x)$, generates $SU(2)$
gauge transformations on the phase space, while the smeared
constraint $\mathcal {V}(\overrightarrow{N}):=\int_\Sigma
d^3xN^a(V_a-A_a^i\mathcal {G}_i)$ generates spatial
diffeomorphism transformations. Together with the
smeared Hamiltonian constraint $H(N)=\int_\Sigma d^3xNH$, the
constraints algebra has the following form\cite{Zh10}: \ba \{\mathcal
{G}(\Lambda),\mathcal {G}(\Lambda^\prime)\}&=&\mathcal
{G}([\Lambda,\Lambda^\prime]), \nn\\
\{\mathcal
{G}(\Lambda),\mathcal{V}(\overrightarrow{N})\}&=&-\mathcal{G}(\mathcal
{L}_{\overrightarrow{N}}\Lambda,),\nn\\
\{\mathcal {G}(\Lambda),H(N)\}&=&0,\nn\\
\{\mathcal {V}(\overrightarrow{N}),\mathcal
{V}(\overrightarrow{N}^\prime)\}&=&\mathcal
{V}([\overrightarrow{N},\overrightarrow{N}^\prime]), \nn\\
\{\mathcal {V}(\overrightarrow{N}),H(M)\}&=&\mathcal{H}(\mathcal
{L}_{\overrightarrow{N}}M),\nn\\
\{H(N),H(M)\}&=&\mathcal {V}(ND^aM-MD^aN). \ea Hence the constraints
are of first class. The total Hamiltonian is a linear combination of
constraints as
\ba \mathcal {H}\equiv\int_\Sigma
H(N)+\mathcal {V}(\overrightarrow{N})+\mathcal {G}(\Lambda).\ea
To
summarize, $f(\R)$ theories of gravity have been cast into the
$\SU(2)$-connection dynamical formalism. Though a scalar field is
non-minimally coupled, the resulted Hamiltonian structure is similar
to GR. Note that what we obtain is real $\SU(2)$-connection dynamics of $f(\R)$ gravity
rather than complex connection dynamics of some conformal theories\cite{Fa10}.

Now the non-perturbative loop quantization procedure can be
straightforwardly extended to $f(\R)$ theories. Since the
configuration space consists of geometry sector  and scalar sector,
we expect the kinematical Hilbert space of the system to be a direct
product of the Hilbert space of geometry and that of scalar field.
To construct quantum kinematics for geometry as in LQG, we have to
extend the space $\mathscr{A}$ of smooth connections to space
$\bar{\mathscr{A}}$ of distributional connections. A simple element
$\bar{A}\in\bar{\mathscr{A}}$ may be thought as a holonomy,
$h_e(A)=\mathcal {P}\exp\int_eA_a$, of a connection along an edge
$e\subset\Sigma$. Through projective techniques, $\bar{\mathscr{A}}$
is equipped with a natural measure $\mu_0$, called the
Ashtekar-Lewandowski measure\cite{As04,Ma07}. In a certain
sense, this measure is the unique diffeomorphism and internal gauge
invariant measure on $\bar{\mathscr{A}}$\cite{Th07}. The kinematical
Hilbert space of geometry then reads
$\hil^\grav_\kin=L^2(\bar{\mathscr{A}},d\mu_0)$. A typical vector
$\Psi_\alpha(\bar{A})\in\hil^\grav_\kin$ is a cylindrical function
over some finite graph $\alpha\subset\Sigma$. The so-called
spin-network basis $T_\alpha(A)\equiv T_{\alpha,j,m,n}(\bar{A})$
provides an orthonormal basis for
$\hil^\grav_\kin$\cite{As04,Ma07}. Note that the spatial
geometric operators of LQG, such as the area, the volume and the
length operators\cite{05} are still valid here. Since the scalar
field also reflects $f(\R)$ gravity, it is natural to employ the
polymer-like representation for it's quantization \cite{06,Ma06}. In
this representation, one extends the space $\mathscr{U}$ of smooth
scalar fields to the quantum configuration space $\bar{\mathscr{U}}$.
A simple element $U\in\bar{\mathscr{U}}$ may be thought as a
point holonomy, $U_\lambda=\exp(i\lambda\phi(x))$, at point
$x\in\Sigma$, where $\lambda$ is a real number. By GNS
structure\cite{Th07}, there is a natural diffeomorphism invariant
measure $d\mu$ on $\bar{\mathscr{U}}$\cite{06}. Thus the kinematical Hilbert
space of scalar field reads
$\hil^\sca_\kin=L^2(\bar{\mathscr{U}},d\mu)$. The following
scalar-network functions of $\phi$, \ba T_X(\phi)\equiv
T_{X,\lambda}(\phi)=\prod_{x_j\in X}U_{\lambda_j}(\phi(x_j)) \ea where
$X=\{x_1,\dots, x_n\}$ is an arbitrary given set of finite number
of points in $\Sigma$, constitute an orthonormal basis in
$\hil^{\sca}_\kin$. Thus the total kinematical Hilbert space for
$f(\R)$ gravity reads $\hil_\kin:=\hil^\grav_\kin\otimes
\hil^\sca_\kin$ with an orthonormal basis
$T_{\alpha,X}(A,\phi)\equiv T_{\alpha}(A)\otimes T_{X}(\phi)$.
A basic feature of loop quantization is that only holonomies
will become configuration operators, rather than the classical
configuration variables themselves. Since the holonomy of a
connection is smeared over an 1-dimensional curve, the conjugate
densitized triad is smeared over 2-surfaces as
$E(S,f):=\int_S\epsilon_{abc}E^a_if^i$, where $f^i$ is a
$\SU(2)$-valued function on $S$. Since the point holonomy of a
scalar is defined on an 0-dimensional point, the momentum is smeared
on 3-dimensional regions $R$ in $\Sigma$ as $\pi(R):=\int_R
d^3x\pi(x)$. Let $\Psi(A,\phi)$ denote a quantum state in
$\hil_\kin$. Then the actions of basic operators read \ba
\hat{h}_e(A)\Psi(A,\phi)&=&h_e(A)\Psi(A,\phi),
\nn\\
\hat{E}(S,f)\Psi(A,\phi)&=&i\hbar\{E(S,f),\Psi(A,\phi)\},\nn\\
\hat{U}_\lambda(\phi(x))\Psi(A,\phi)&=&\exp(i\lambda\phi(x))\Psi(A,\phi), \nn\\
\hat{\pi}(R)\Psi(A,\phi)&=&i\hbar\{\pi(R),\Psi(A,\phi)\}.  \ea As in
LQG, it is straightforward to promote the
Gaussian constraint $\mathcal {G}(\Lambda)$ to a well-defined
operator in $\hil_\kin$\cite{Ma07}. It's kernel is the internal gauge invariant
Hilbert space $\mathcal {H}_G$ with gauge invariant spin-network
basis as well. Since the diffeomorphisms of $\Sigma$ act covariantly
on the cylindrical functions in $\mathcal {H}_G$, the so-called
group averaging technique can be employed to solve the
diffeomorphism constraint\cite{As04,Ma07}. Thus we can also obtain the desired
diffeomorphism and gauge invariant Hilbert space $\mathcal
{H}_{Diff}$ for $f(\R)$ gravity.

The nontrivial task is to implement the Hamiltonian constraint
$H(N)$ at quantum level. As in LQG, we can show by detail and
technical analysis that the Hamiltonian constraint can be promoted
to a well-defined operator in $\hil_\kin$\cite{Zh10}. The resulted Hamiltonian constraint operator
is internal gauge invariant and diffeomorphism covariant. Hence it
is at least also well defined in $\mathcal {H}_G$. Comparing Eq.(\ref{hamilton}) with the Hamiltonian
constraint of GR in connection formalism\cite{Th07}, the new
ingredients of $f(\R)$ gravity that we have to deal with are
$\phi(x),\phi^{-1}(x),V(\phi)$ and the following four terms \ba
H_3&=&\int_\Sigma d^3x\frac
N{3\phi}\frac{(\tilde{K}^i_aE^a_i)^2}{\sqrt{h}},\nn\\
H_4&=&\int_\Sigma d^3x \frac{2N}{3}
\frac{(\tilde{K}^i_aE^a_i)\pi}{\sqrt{h}},\nn\\
H_6&=&\int_\Sigma d^3x
\frac{N}{3}\frac{\pi^2\phi}{\sqrt{h}},\nn\\
H_7&=&\int_\Sigma d^3x N\sqrt{h}D_aD^a\phi. \ea By introducing
certain small constant $\lambda_0$, an operator corresponding to the
scalar $\phi(x)$ can be defined as \ba
\hat{\phi}(x)=\frac{1}{2i\lambda_0}(U_{\lambda_0}(\phi(x))-U_{-\lambda_0}(\phi(x))).
\ea The ambiguity of $\lambda_0$ is the price that we have to pay in
order to represent field $\phi$ in the polymer-like representation.
To further define an operator corresponding to $\phi^{-1}(x)$, we
can use the classical identity \ba
\phi^{-1}(x)=(\frac1l\{\phi^{l}(x),\pi(R)\})^{\frac{1}{1-l}},
 \ea for any rational number $l\in(0,1)$. For example, one may choose $l=\frac12$ for positive $\phi(x)$ and
replace the Poisson bracket by commutator to define \ba
\hat{\phi}^{-1}(x)=(\frac{2}{i\hbar}[\sqrt{\hat{\phi}(x)},\hat{\pi}(R)])^2.
\ea Similar tricks can be employed to deal with the function
$V(\phi)$, provided that it can be expanded as powers of $\phi(x)$.
Moreover, by the regularization techniques developed for the
Hamiltonian constraint operators of LQG\cite{Th07} and polymer-like
scalar field\cite{Ma06}, all the terms $H_3,H_4,H_6$ and $H_7$ can
be quantized as operators acting on cylindrical functions in
$\hil_\kin$ in state-dependent ways\cite{Zh10}. For example, the
operator corresponding to $H_4$ acts on a basis vector as \ba
\hat{H}_4\cdot T_{\alpha,X} &=&\sum_{v\in
v(\alpha)}\frac{2^{14}N(v)}{3^4\gamma^4(i\hbar)^6}C(v)\hat{\pi}_v
\hat{h}^{-1}_{s_L(v)}[\hat{h}_{s_L(v)},\hat{\kt}]
\nn\\
&\times&\epsilon^{LMN}\Tr(\hat{h}^{-1}_{s_M(v)}[\hat{h}_{s_M(v)},(\hat{V}_{v})^{3/4}]\nn\\
&\times&\hat{h}^{-1}_{s_N(v)}[\hat{h}_{s_N(v)},(\hat{V}_{v})^{3/4}]) \nn\\
&\times&\epsilon^{IJK}\Tr(\hat{h}^{-1}_{s_I(v)}[\hat{h}_{s_I(v)},(\hat{V}_{v})^{1/2}]\nn\\
&\times&
\hat{h}^{-1}_{s_J(v)}[\hat{h}_{s_J(v)},(\hat{V}_{v})^{1/2}] \nn\\
&\times&\hat{h}^{-1}_{s_K(v)}[\hat{h}_{s_K(v)},(\hat{V}_{v})^{1/2}])\cdot
T_{\alpha,X} \label{T4}\ea where the coefficient $C(v)$ comes from the triangulation
ambiguity, and $h_{s_I(v)}$ denotes the holonomy along the segment $s_I$
starting from the vertex $v$ of graph $\alpha$. Note that the action
of the volume operator $\hat{V}$ on
$T_\alpha(A)$ over a graph $\alpha$ can be factorized as
$\hat{V}\cdot T_\alpha=\sum_{v\in V(\alpha)}\hat{V}_v\cdot
T_\alpha$. The action of the operator $\pi(R)$ on $T_X(\phi)$ over a graph $X$ can also be factorized as
$\hat{\pi}(R)\cdot T_X=\sum_{x_i\in X\cap R}\hat{\pi}_{x_i}\cdot
T_X$. It is easy to see from Eq.(\ref{T4}) that the action of $\hat{H}_4$ on $
T_{\alpha,X}$ is graph changing. It adds a finite number of vertices
within the edges $e_I(t)$ starting from each high-valence
vertex of $\alpha$. By similar ways, the whole Hamiltonian constraint can be
quantized as a well-defined operator $\hat{H}$, which is internal gauge invariant and diffeomorphism covariant. Although $\hat{H}$ can dually act on the diffeomorphism
invariant states, there is no guarantee for the resulted states to
be still diffeomorphism invariant. Hence it is difficult to define a
Hamiltonian constraint operator directly in $\hil_{Diff}$. One way
out is to employ the master constraint program\cite{07,08}. By
using the structure of $\hat{H}$, we can define also a corresponding master constraint
operator in $\hil_{Diff}$\cite{Zh10}. Then it is very possible to
solve all the quantum constraints and obtain some physical Hilbert
space with observables in it.

We summarize with a few remarks. (i) The connection dynamics of
$f(\R)$ gravity has been obtained by canonical transformations from
it's geometric dynamics. It is still desirable to find an action for
the connection dynamics. (ii) Due to the $\SU(2)$-connection
dynamical formalism, the metric
$f(\R)$ theories have been successfully quantized by extending LQG scheme. Thus, the non-perturbative loop quantization
procedure is not only valid for GR but also valid for a rather
general class of 4-dimensional metric theories of gravity. (iii)
Classically the scalar fields $\phi$ characterize different $f(\R)$
theories of gravity by $\phi=f'(\R)$. Thus for a given $f(\R)$
theory, $\phi$ will become a particular function of $\R$ while the potential $V(\phi)$ is fixed. Hence our quantum $f(\R)$ gravity may be understood as a class
of quantum theories representing different choices of the function
$f(\R)$. However, the other possible and appealing interpretation remains. We may
just think different classical $f(\R)$ theories as emerging from
 different classical limits of the quantum observables
$\hat{\phi}$ and $\hat{\R}$. The latter understanding provides an enlightening
mechanism to produce chameleon $f(\R)$ theories from one fundamental quantum gravity theory, which might be
significant to understand our universe.

We would like to thank Nathalie Deruelle and  Yuuiti Sendouda for
helpful discussion. This work is supported by NSFC (No.10975017) and
the Fundamental Research Funds for the central Universities.


\newcommand\AL[3]{~Astron. Lett.{\bf ~#1}, #2~ (#3)}
\newcommand\AP[3]{~Astropart. Phys.{\bf ~#1}, #2~ (#3)}
\newcommand\AJ[3]{~Astron. J.{\bf ~#1}, #2~(#3)}
\newcommand\APJ[3]{~Astrophys. J.{\bf ~#1}, #2~ (#3)}
\newcommand\APJL[3]{~Astrophys. J. Lett. {\bf ~#1}, L#2~(#3)}
\newcommand\APJS[3]{~Astrophys. J. Suppl. Ser.{\bf ~#1}, #2~(#3)}
\newcommand\JCAP[3]{~JCAP. {\bf ~#1}, #2~ (#3)}
\newcommand\LRR[3]{~Living Rev. Relativity. {\bf ~#1}, #2~ (#3)}
\newcommand\MNRAS[3]{~Mon. Not. R. Astron. Soc.{\bf ~#1}, #2~(#3)}
\newcommand\MNRASL[3]{~Mon. Not. R. Astron. Soc.{\bf ~#1}, L#2~(#3)}
\newcommand\NPB[3]{~Nucl. Phys. B{\bf ~#1}, #2~(#3)}
\newcommand\PLB[3]{~Phys. Lett. B{\bf ~#1}, #2~(#3)}
\newcommand\PRL[3]{~Phys. Rev. Lett.{\bf ~#1}, #2~(#3)}
\newcommand\PR[3]{~Phys. Rep.{\bf ~#1}, #2~(#3)}
\newcommand\PRD[3]{~Phys. Rev. D{\bf ~#1}, #2~(#3)}
\newcommand\SJNP[3]{~Sov. J. Nucl. Phys.{\bf ~#1}, #2~(#3)}
\newcommand\ZPC[3]{~Z. Phys. C{\bf ~#1}, #2~(#3)}
\newcommand\CQG[3]{~Class. Quant. Grav. {\bf ~#1}, #2~(#3)}


\begin{thebibliography}{}

\bibitem{Ro04} C. Rovelli, Quantum Gravity, (Cambridge University Press, 2004).

\bibitem{Th07} T. Thiemann, Modern Canonical Quantum General Relativity, (Cambridge University
Press, 2007).

\bibitem{As04}A. Ashtekar and J. Lewandowski, Class. Quant. Grav. {\bf21}, R53 (2004).

\bibitem{Ma07} M. Han, W. Huang, and Y. Ma,
Int. J. Mod. Phys. D {\bf16}, 1397 ,(2007).

\bibitem{Le10}M. Domagala, K. Giesel, W. Kaminski, J. Lewandowski, Phys. Rev. D {\bf 82}, 104038 (2010).

\bibitem{As86}A. Ashtekar,  Phys. Rev. Lett. {\bf
57}, 2244 (1986).

\bibitem{Ba}J. Barbero,  Phys. Rev. D {\bf
51}, 5507 (1995).

\bibitem{01}J. Friemann, M. Turner, D. Huterer, Ann. Rev. Astron.
Astrophys. {\bf46}, 385 (2008).


\bibitem{So} T. P. Sotiriou, V. Faraoni,
 Rev. Mod. Phys. {\bf82}, 451 (2010).

\bibitem{Na}N. Deruelle, Y. Sendouda, and A. Youssef,
 Phys. Rev. D {\bf 80}, 084032 (2009).

\bibitem{Na01} N. Deruelle, M. Sasaki, Y. Sendouda, D.
Yamauchi, Prog. Theor. Phys. {\bf123}, 169 (2010).

\bibitem{Zh10}The details will appear in a succeeding article by X. Zhang and Y.
Ma.

\bibitem{Fa10} L. Fatibene, M. Ferraris, M. Francaviglia,  Class. Quant. Grav. {\bf27}, 185016 (2010).

\bibitem{05}Y. Ma, C. Soo, J. Yang, Phys. Rev. D {\bf81},
124026 (2010).

\bibitem{06}A. Ashtekar, J. Lewandowski, H. Sahlmann, Class. Quant. Grav. {\bf20}, L11 (2003).

\bibitem{Ma06} M. Han and Y. Ma,  Class. Quant. Grav. {\bf23}, 2741 (2006).

\bibitem{07}T. Thiemann, Class. Quant. Grav. {\bf23}, 3211 (2006).

\bibitem{08}M. Han and Y. Ma, Phys. Lett. B {\bf634}, 225 (2006).


\end{thebibliography}
\end{document}